\documentclass[cits]{PoS}
\usepackage[utf8]{inputenc}
\usepackage{graphicx}
\usepackage[usenames]{color}
\let\normalcolor\relax
\usepackage{amssymb}
\usepackage{amsmath}
\usepackage{nicefrac}
\usepackage{multirow}
\usepackage{subfig}
\usepackage{dcolumn}

\newcommand{\sh}[1]{#1\hskip-7pt \diagup}

\title{Model comparison of Delta and Omega masses in a covariant Faddeev approach}

\ShortTitle{Model comparison of Delta and Omega masses in a covariant Faddeev approach}

\author{\speaker{Helios Sanchis-Alepuz}\\
        Institut f\"ur Physik, Karl-Franzens--Universit\"at Graz, Universit\"atsplatz 5, 8010 Graz, Austria\\
        E-mail: \email{helios.sanchis-alepuz@uni-graz.at}}

\author{Reinhard Alkofer\\
        Institut f\"ur Physik, Karl-Franzens--Universit\"at Graz, Universit\"atsplatz 5, 8010 Graz, Austria\\
        E-mail: \email{reinhard.alkofer@uni-graz.at}}

\author{Gernot Eichmann\\
        Institut f\"ur Theoretische Physik, Justus-Liebig-Universit\"at
	  Giessen, D-35392 Giessen, Germany\\
        E-mail: \email{gernot.eichmann@theo.physik.uni-giessen.de}}

\author{Richard Williams\\
        Departamento de F\'{\i}sica Te\'orica I,  Universidad Complutense, 28040 Madrid, Spain\\
        E-mail: \email{richard.williams@fis.ucm.es}}

\abstract{We compute the vector-meson, nucleon and $\Delta / \Omega$-baryon 
masses and their evolution with the current-quark mass
using a covariant generalized Bethe-Salpeter equation approach. The
interaction kernel is truncated to a dressed gluon exchange. We study
the model dependence of our results with the quark-gluon dressing to
assess the validity of the truncation.}

\FullConference{International Workshop on QCD Green's Functions, Confinement and Phenomenology\\
		 5-9 September 2011\\
		 Trento, Italy}

\begin{document}

\section{Introduction}
The covariant approach to bound-state calculations provides a very
powerful tool for the study of relativistic two- and three-body bound
states. When applied to hadrons, they rely upon the knowledge of QCD
Green's functions. The intrinsic sophistication of these equations,
however, presents several difficulties in practical applications due
to the complexity of taking such calculations beyond the simplest
truncation known as Rainbow-Ladder (RL).

It is only in recent years that a unified approach has been reached for the
study of mesons and baryons, with quark-diquark calculations
\cite{hep-ph/9705267,nucl-th/9805054,nucl-th/9907120,arXiv:0812.1665,arXiv:1008.3184}
surpassed by their more intricate three-body description
\cite{arXiv:0912.2246,arXiv:1104.4505,arXiv:1109.0199}. At
the same time however, significant technical progress has
been made in the covariant treatment of mesons beyond the RL
truncation \cite{arXiv:0808.3372,arXiv:0905.2291,arXiv:0903.5461}. In
the case of baryons, these calculations are, due to technical
difficulties, so far restricted to particles composed of quarks of the
same mass. Therefore it is not feasible for the moment to calculate the
plethora of baryon masses, as is done in other approaches such as
constituent quark models (see, e.g. \cite{arXiv:0811.1752} and
references therein). Nevertheless, one of the main goals of the approach
must be to identify the
relevance of the different quantum-field-theoretical interaction terms
in bound-state phenomena, and in this respect it represents an excellent
tool that is complementary to lattice QCD.

So far, the calculation requires to model the quark-gluon interaction
via an effective interaction. However, it is interesting to note that
studies within RL
have been dominated by just one effective interaction, known as the Maris-Tandy model~\cite{nucl-th/9708029,nucl-th/9905056}.
On the one hand, this dominance is well-earned
since this ansatz performs very well indeed. On
the other hand, with the swift improvement in our knowledge of QCD Green's
functions from both lattice and functional approaches, it is possible to
define different effective interactions which, presumably, capture more
faithfully some of QCD's features. Note that this is not to say that
they will perform better \emph{phenomenologically}.
Therefore, before the daunting challenge of bringing covariant baryon
studies beyond RL (such as the inclusion of pion-cloud
effects and three-body forces) or drawing conclusions about the role of
these different contributions, one should thoroughly investigate the
model-independent features within a given truncation scheme.

\section{Framework}

\subsection{Bound-state equations}
The covariant description of bound-state equations begins with Dyson's
equation for the connected and amputated $n$-quark scattering matrix $T$:
\begin{equation}
	T = K + K G_0 T \,\,,
\end{equation}
where $G_0$ is the disconnected $n$-quark dressed propagator and $K$ is the
$n$-body kernel that details the interaction between the quarks that
constitute the bound state.

It is this interaction kernel $K$ that proves to be
challenging. For mesons we must respect
chiral symmetry in order to realise the pion as a (pseudo)-Goldstone boson. This gives a precise relationship between the truncation
of the quark-propagator Dyson-Schwinger equation (DSE) and the truncation of the 2-body kernel
here. The simplest example of a symmetry-preserving kernel is that of
Rainbow-Ladder (RL), though more sophisticated kernels have been
studied~\cite{arXiv:0808.3372,arXiv:0905.2291,arXiv:0903.5461}.

\subsection{Quark propagator and DSE}
The first ingredient for covariant bound-state studies is the dressed quark propagator,
\begin{equation}
	S^{-1}(p) = A(p^2) \left(  i\sh{p} + M(p^2) \right)\,\,,
	\label{eqn:inverse_quark_propagator}
\end{equation}
where $1/A(p^2)$ is the quark wave-function renormalisation and $M(p^2)$
is the quark mass function. These scalar dressing functions are obtained
as solutions to the quark DSE,
\begin{equation}
	S^{-1}(p) = Z_2 \,S_0^{-1} + g^2 \,Z_{1f}\int \frac{d^4k}{\left( 2\pi
	\right)^4} \gamma^\mu S(k) \,\Gamma^\nu(k,p) \,D_{\mu\nu}(q)\,\,,
	\label{eqn:quark_DSE}
\end{equation}
where $q=k-p$ is the momentum of the exchanged gluon. Here $S_0^{-1}(p)$ represents the
bare inverse quark propagator, obtained
from Eq.~(\ref{eqn:inverse_quark_propagator}) by setting $A(p^2)=1$ and
$M(p^2)=m_0$. This bare mass is related to the renormalised one
via $Z_2 m_0 = Z_2 Z_m m_q$, with $Z_2$ and $Z_m$ the wave-function and
quark-mass renormalisation constants, and $Z_{1f}$ that of the quark-gluon vertex.
The cardinal input in the quark propagator DSE~(\ref{eqn:quark_DSE}) are the gluon propagator $D_{\mu\nu}(q)$ and
the quark-gluon vertex $\Gamma^\nu(k,p)$.

\subsection{Rainbow-Ladder}
The quark-gluon interaction that appears in DSE for the quark reads:
\begin{equation}\label{DSEkernel}
  Z_{1f}\,\frac{g^2}{4\pi}\,D_{\mu\nu}(q)\,\Gamma_\nu(k,p) \,.
\end{equation}
In Landau gauge, $D_{\mu\nu}$ is just the transverse projector
$T_{\mu\nu}(q) = \delta_{\mu\nu} - q_\mu q_\nu/q^2$ multiplied by the
scalar gluon dressing function $Z(q^2)/q^2$. The quark-gluon vertex $\Gamma_\nu(k,p)$ is
decomposed as twelve Dirac covariants, of which the minimal set in
Landau gauge numbers eight. The full vertex can be written
as the sum of its bare tree-level part plus a self-energy correction: $\Gamma_\nu(k,p) = Z_{1f}\gamma_\nu + \Lambda_\nu$.

The RL truncation requires that we replace the complicated structure of the quark-gluon
vertex with the $\gamma_\mu$ projection of the non-perturbative
corrections. Hence Eq.~(\ref{DSEkernel}) becomes
\begin{equation}\label{eqn:trunc2}
Z_{1f}\,\frac{g^2}{4\pi}\,
T_{\mu\nu}(q) \,\frac{Z(q^2)}{q^2}\,\left( Z_{1f} + \Lambda(q^2) \right)
\gamma_\nu\,\,,
\end{equation}
where now $\Lambda(q^2)$ is the non-perturbative dressing of the
$\gamma_\nu$ part of the quark-gluon vertex, restricted to depend only
on the exchanged gluon momentum. The remaining structures are as
before.
If one wishes to draw a distinction between the gluon dressing $Z(q^2)$
and the quark-gluon dressings, it can be useful to define the function
$\Gamma_{YM}(q^2) \equiv Z_{1f} + \Lambda(q^2)$. This 
distinction is very important because the gluon propagator is by now
well-known from both Lattice studies and other functional approaches.
However, if one wishes to take a purely phenomenological approach one
can instead combine all scalar dressings into one effective running
coupling, $\alpha_{eff}(q^2)$:
\begin{equation}
Z_{1f}\,\frac{g^2}{4\pi} \,D_{\mu\nu}(q) \,\Gamma_\nu(k,p)
  = Z_2^2 \, T_{\mu\nu}(q) \,\frac{\alpha_{eff}(q^2)}{q^2}\,\gamma_\nu\,.
\end{equation}
In all cases, the Dirac structure remains the same, and $Z_2^2$ follows from the
Slavnov-Taylor identities to maintain multiplicative renormalisability.
When it comes to discussing the possible impact of effects beyond RL, it
is convenient to think in terms of the separate dressing functions $Z$ and
$\Lambda$.

The symmetry-preserving two-body kernel corresponding to this yields
the `ladder' part of RL. For simplicity we quote this in terms of
$\alpha_{eff}$,
\begin{equation}
	K^{\textrm{2-body}}= 4\pi \,Z_2^2 \,\frac{\alpha_{eff}(q^2)}{q^2}\,
	T_{\mu\nu}(q)\,\gamma^\mu \otimes \gamma^\nu\,\,.
	\label{eqn:ladder}
\end{equation}
For the baryon, the three-body kernel $K^{\textrm{3-body}}$ is decomposed into a three-quark
irreducible contribution $K^{\textrm{3-body}}_{\textrm{irr}}$ and the sum of permuted two-body
kernels $K^{\textrm{2-body}}_{(a)}$, with the subscript $a$ indicating the spectator quark:
\begin{equation}
	K^{\textrm{3-body}} = K^{\textrm{3-body}}_{\textrm{irr}} +
	\sum_{a=1}^{3} S_{(a)}^{-1} \otimes K^{\textrm{2-body}}_{(a)}\,\,.
	\label{eqn:three-body}
\end{equation}
Note that for mesons the two-body kernel describes quark-antiquark
correlations, whilst in the baryon it pertains to quark-quark
correlations.  Motivated by the success of quark-diquark
calculations, we ignore $K^{\textrm{3-body}}_{\textrm{irr}}$.

\subsection{Model interaction}
When one constructs a model interaction our first constraint is that of
perturbation theory. This only fixes the large momentum behaviour,
leaving the IR behaviour unspecified. This is typically chosen such that Dynamical
Chiral Symmetry Breaking (DCSB) is realised. The chiral condensate,
or equivalently the pion decay constant, determines
the strength of chiral-symmetry breaking. These are fixed to
their phenomenological/experimental values, respectively.  With a symmetry
preserving truncation protecting the chiral behaviour, such chiral
properties are guaranteed and the characteristic $m_\pi \propto m_q^{1/2}$
behaviour is seen for `small' $m_q$.

The vector meson mass is similarly determined, to a large degree, by the
strength of breaking of chiral symmetry, albeit linearly with respect
to the quark mass. That the interaction thus far constructed also works
well here is not surprising, since the vector mesons are $1S$ states
and similarly insensitive to $L.S$ couplings present in the
$\gamma^\mu \otimes \gamma^\mu$ interaction, that in RL are too attractive.
Of course, meson bound-states with different quantum numbers are not so
easily described. This is particularly true for the ground-state axial
vectors which are significantly overbound here.

In the present work we employ two model interactions.
In the Maris-Tandy (MT) model~\cite{nucl-th/9708029,nucl-th/9905056} the effective running
coupling is given by
	\begin{eqnarray}
	\alpha_{eff}(q^2) &=&
      	 \pi\eta^7\,x^2 \, e^{-\eta^2\,x}+\frac{2\pi\gamma_m \big(1-e^{-y}\big)}{\log\,[e^2-1+(1+z)^2]}\,, \quad
          \begin{array}{rl}
       x &= q^2/\Lambda^2\,, \\
       y &= q^2/\Lambda_{t}^2\,, \\
       z &= q^2/\Lambda_{QCD}^2\,,
    \end{array}
	\end{eqnarray}
and features a Gaussian distribution in the infrared that provides dynamical chiral symmetry breaking.
It is characterized by an energy scale $\Lambda=0.74$ GeV, fixed to give the pion decay constant, and a dimensionless parameter $\eta$.
Many ground-state hadron observables have been found to be almost insensitive to the value of $\eta$ around $\eta=1.8$.
The second part reproduces the one-loop running coupling at large, perturbative, momenta.
It includes the anomalous dimension $\gamma_m=12/(11N_C-2N_f)$ of the quark propagator, and
we use $\gamma_m=12/25$, $\Lambda_{QCD}=0.234$ GeV and $\Lambda_t=1$ GeV.
Note that we also employ a Pauli-Villars like regulator with a mass scale of $200$~GeV.
The quark masses at $\mu=19$~GeV are $3.7$, $85.2$, $869$ and $3750$ MeV for the
$u/d$, $s$, $c$, and $b$ quarks, respectively.

The Alkofer-Fischer-Williams (AFW) model~\cite{arXiv:0804.3478}, on the other hand, is motivated by the desire to account
for the $U_A(1)$-anomaly by the Kogut-Susskind
mechanism. The effective coupling is
constructed as the product of the gluon dressing~\cite{hep-ph/0309077}
 and a model for
the non-perturbative behaviour of the quark-gluon
vertex~\cite{arXiv:0804.3042},
\begin{equation}
  \alpha_{eff}(q^2) =  \mathcal{C} \left(\frac{x}{1+x}\right)^{2\kappa}
  \left(\frac{y}{1+y}\right)^{-\kappa-1/2}
  \left( \frac{\alpha_0+a_{UV}\,x}{1+x} \right)^{-\gamma_0}
  \left( \lambda +\frac{a_{UV}\,x}{1+x} \right)^{-2\delta_0}\,\,.
\end{equation}
The four terms in parentheses are: the IR scaling of the gluon
propagator; IR scaling of the quark-gluon vertex; logarithmic running of
the gluon propagator; and the logarithmic running of the quark-gluon
vertex. Additionally, the last two are constructed to interpolate between the IR and UV
behaviour. The remaining terms are defined as follows:
\begin{equation}
    \lambda = \frac{\lambda_S}{1+y} + \frac{\lambda_B \,y}{1+(y-1)^2}\,, \quad
    a_{UV}= \pi  \gamma_m \left( \frac{1}{\ln{z}}-\frac{1}{z-1} \right), \quad
    \begin{array}{rl}
       x &= q^2/\Lambda_{YM}^2\,, \\
       y &= q^2/\Lambda_{IR}^2\,, \\
       z &= q^2/\Lambda_{UV}^2\,,
    \end{array}
\end{equation}
and $\alpha_0=8.915/N_C$. Here, $\Lambda_{YM}= 0.71$ GeV is the
dynamically generated Yang-Mills scale, while $\Lambda_{UV}= 0.5$ GeV corresponds
to the one-loop perturbative running. The IR scaling exponent is
$\kappa=0.595353$, and the one-loop anomalous dimensions are
related via $1+\gamma_0 = -2\delta_0 = \frac{3}{8}\,N_C \,\gamma_m$, with $\gamma_m=12/(11N_C-2N_f)$.
We choose $N_f=5$ active quark flavours at the
renormalisation point $\mu=19$ GeV. The constant
$\mathcal{C}=0.968$ is chosen such that $\alpha_{eff}$ runs appropriately in
the UV. Finally, $\Lambda_{IR}=0.42$ GeV, $\lambda_S=6.25$, and $\lambda_{B}=21.83$
determine the IR properties of the quark-gluon vertex and are fitted such that the properties of $\pi$, $K$ and $\rho$
mesons are well reproduced.  The quark
masses at $\mu=19$~GeV are $2.76$, $55.3$, $688$ and $3410$ MeV for the
$u/d$, $s$, $c$, and $b$ quarks, respectively.

\section{Results and discussion}

\begin{table*}[p]
\begin{center}
\begin{tabular}{|c@{\!\;\;}l|ccc|} \hline
\multicolumn{2}{|c|}{$J^{PC}=0^{-+}$}  & MT         & AFW         & exp. \\ \hline
$n\overline{n}$ &$(\pi)$               & 0.140\dag  & 0.139\dag   & 0.138 \\
$n\overline{s}$ &$(K)$                 & 0.496\dag  & 0.497\dag   & 0.496 \\
$s\overline{s}$ &                      & 0.697      & 0.686       & -- \\
$c\overline{c}$ &$(\eta_c)$            & 2.979\dag  & 2.980\dag   & 2.980 \\
$b\overline{b}$ &$(\eta_b)$            & 9.388\dag  & 9.390\dag   & 9.391 \\ \hline
\multicolumn{2}{|c|}{$J^{PC}=1^{--}$}  & MT         & AFW         & exp. \\ \hline
$n\overline{n}$ &$(\rho)$              & 0.743      & 0.710       & 0.775 \\
$n\overline{s}$ &$(K^\star)$           & 0.942      & 0.961       & 0.892 \\
$s\overline{s}$ &$(\phi)$              & 1.075      & 1.114       & 1.020 \\
$c\overline{c}$ &$(J/\psi)$            & 3.163      & 3.302       & 3.097 \\
$b\overline{b}$ &$(\Upsilon)$          & 9.466      & 9.621       & 9.460 \\ \hline
\end{tabular} \hspace{5mm}
\begin{tabular}{|c|cccc|} \hline
                    & MT    & AFW   & exp. &   \\ \hline
$N$                 & 0.94  & 0.97  & 0.94 &      \\
$\Delta$            & 1.26  & 1.22  & 1.23 &      \\
$\Omega$            & 1.72  & 1.80  & 1.67 &       \\ \hline
                    & MT    & AFW   & lattice  & LPW  \\ \hline
$\Omega_{ccc}$      & 4.4   & 4.9   & 4.7      & 4.9(0.25)        \\
$\Omega_{bbb}$      & 13.7  & 13.8  & 14.4     & 14.5(0.25)        \\
\hline
\end{tabular}
\caption{Computed meson and baryon masses (GeV) for both MT and AFW
interactions, compared to experiment. Quantities fitted to their experimental values are
indicated by a $\dag$. Since the heavy-Omega baryons have not been observed yet, we compare
to lattice calculations~\cite{hep-lat/0501021,arXiv:1010.0889}
 and a recent study from pNRQCD~\cite{arXiv:1111.7087}.
}\label{tab:results}
\end{center}
\end{table*}

\begin{figure}[p]
\centering
\includegraphics[height=0.48\textheight,clip]{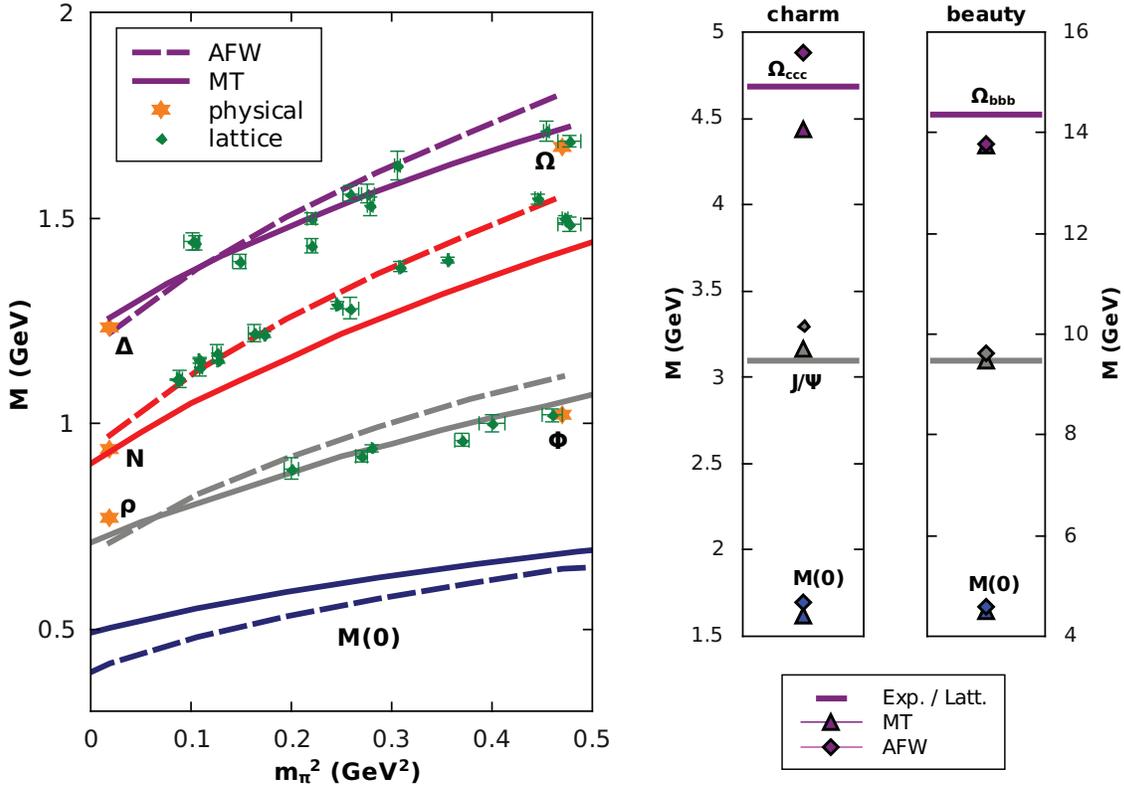}
\caption{\textit{Left panel:} evolution of $M(0)$ as well as $\rho$, $N$ and $\Delta$ masses
with the squared pion mass for MT and AFW models. Results are compared
to lattice data; see~\cite{arXiv:1104.4505,arXiv:1109.0199} for references.
\textit{Right panel:} vector-meson and triply-heavy omega masses for the two interaction models
MT and AFW.}
\label{fig:figs}
\end{figure}

Using the techniques described in~\cite{arXiv:1104.4505,arXiv:1109.0199},
we calculated the vector-meson, nucleon, and delta\footnote{Assuming isospin symmetry, $\Delta$ and $\Omega$ baryons have the same structure, changing only the current-quark mass.} masses
up to the bottom region using the AFW and MT effective interactions.
The results are collected in Table~\ref{tab:results}, and their evolution
with the pseudoscalar mass (or, equivalently, with the current-quark mass)
is shown in Fig.~\ref{fig:figs}.

Studying different interactions allows to quantify the model dependence within the rainbow-ladder truncation.
We find that the two interaction models yield similar overall trends in the results.
There is little deviation in the resulting hadron masses in the light-quark region, whereas
the MT results tend to underestimate the AFW values for heavier quark masses.
In all cases, the agreement with experimental and lattice data is $\lesssim 10\%$ and
comparable to the frequently studied model dependence in the MT interaction that is induced by the parameter $\eta$.

We note that both AFW and MT interactions, which were designed in a different spirit,
show a qualitatively similar behavior in the mid-momentum range around $|q| \sim 0.5 \dots 1$~GeV
which is the relevant domain for dynamical chiral symmetry breaking. It is this region that
provides the overall strength to ground-state hadron properties, whereas their features are
less sensitive to the deep-infrared region. The same observation has also been made in recent
studies exploring the impact of different model interactions in the light-meson sector~\cite{arXiv:1007.3901,arXiv:1108.0603}.

On the other hand, the impact of DCSB is reduced in the heavy-quark domain
where hadron properties become increasingly sensitive to the heavy quark mass.
We have included the value of the quark mass function $M(p^2=0)$ in both panels of Fig.~\ref{fig:figs}.
Since in the heavy-quark limit the variation of $M(p^2)$ at low momentum
is small, its value is indicative of the heavy-quark constituent mass.
Indeed we find that the spread between the AFW and MT results in the vector-meson and delta channels in the charm and bottom regions
follows the same pattern as that of $M(0)$, indicating that their properties are dominated by the features of the quark DSE
rather than the details of the effective interaction and the structure of the $qq$ kernel that enters the bound-state equations.

In that respect it is interesting to speculate about possible effects beyond the RL truncation.
Such contributions refer to corrections in the quark-gluon vertex
as well as additional structures beyond the vector-vector interaction.
Amongst others, they may consist of attractive pion-cloud corrections in the chiral regime, or
repulsive corrections from self-interactions of the gluon.
They have a two-fold effect: firstly, they allow
for non vector-vector interactions and a more general momentum
dependence; and secondly, they introduce a quark-mass dependence in the
interaction due to the internal quark propagators that are coupled there.
This last point is certainly of relevance when trying to describe such a
wide range of mass scales provided by the $u/d$ to $b$ quarks.

In rainbow-ladder, vertex corrections beyond $\gamma_\nu$ are included in the
modelling; otherwise Dynamical Chiral Symmetry Breaking would be absent.
Since this is done for the light-meson sector, it implies
also that only light quark-dynamics are included. As a result, to
determine how much room there is for beyond-RL effects, one must consider
removing those vertex corrections that are implicitly included. Likely, this can
only be achieved for the heavy-quark sector where one expects vertex
corrections to be suppressed. This will be considered in a future work.

\section{Summary and conclusions}

We have studied the quark-mass dependence of several meson and baryon masses
in the RL truncated Dyson-Schwinger approach.
We investigated a broad range of current-quark masses from the light-quark domain up to the bottom region.
To identify model-independent features,
we employed two different effective interactions for the quark-(anti)quark kernel.
We find that both models yield comparable results that agree with
experimental and lattice data within $\lesssim 10\%$ throughout the quark-mass range, thereby
demarcating the model sensitivity within a RL truncation.

The impact of beyond-RL corrections, on the other hand, is easier to access
in form-factor studies where they are needed for a correct description in the chiral and low-momentum region.
In that respect it is desirable to perform a model comparison of nucleon and delta
electromagnetic form factors in the covariant three-body framework. Work in this direction is in progress.

\acknowledgments
We thank Christian Fischer, Felipe Llanes-Estrada and Selym
Villalba-Chavez for
helpful discussions. This work was supported by  the Austrian Science Fund FWF
under Projects No.\ P20592-N16,  Erwin-Schr\"odinger-Stipendium J3039,
and the Doctoral Program W1203; Ministerio de Educaci\'on No.\ SB2010-0012;
as well as  in part by the European Union
(HadronPhysics2 project ``Study of strongly-interacting matter'').


\begin{thebibliography}{99}

\bibitem{hep-ph/9705267}
  G.~Hellstern, R.~Alkofer, M.~Oettel and H.~Reinhardt,
  Nucl.\ Phys.\ A\ {\bf 627} (1997) 679
  [{\tt hep-ph/9705267}].

\bibitem{nucl-th/9805054}
  M.~Oettel, G.~Hellstern, R.~Alkofer and H.~Reinhardt,
  Phys.\ Rev.\ C\ {\bf 58} (1998) 2459
  [{\tt nucl-th/9805054}].

\bibitem{nucl-th/9907120}
  J.~C.~R.~Bloch, C.~D.~Roberts, S.~M.~Schmidt, A.~Bender and M.~R.~Frank,
  Phys.\ Rev.\ C\ {\bf 60} (1999) 062201
  [{\tt nucl-th/9907120}].

\bibitem{arXiv:0812.1665}
  D.~Nicmorus, G.~Eichmann, A.~Krassnigg and R.~Alkofer,
  Phys.\ Rev.\ D\ {\bf 80} (2009) 054028
  [{\tt arXiv:0812.1665 [hep-ph]}].

\bibitem{arXiv:1008.3184}
  D.~Nicmorus, G.~Eichmann and R.~Alkofer,
  Phys.\ Rev.\ D\ {\bf 82} (2010) 114017
  [{\tt arXiv:1008.3184 [hep-ph]}].

\bibitem{arXiv:0912.2246}
  G.~Eichmann, R.~Alkofer, A.~Krassnigg and D.~Nicmorus,
  Phys.\ Rev.\ Lett.\ \ {\bf 104} (2010) 201601
  [{\tt arXiv:0912.2246 [hep-ph]}].

\bibitem{arXiv:1104.4505}
  G.~Eichmann,
  Phys.\ Rev.\ D\ {\bf 84} (2011) 014014
  [{\tt arXiv:1104.4505 [hep-ph]}].

\bibitem{arXiv:1109.0199}
  H.~Sanchis-Alepuz, G.~Eichmann, S.~Villalba-Chavez and R.~Alkofer,
  Phys.\ Rev.\ D\ {\bf 84} (2011) 096003
  [{\tt arXiv:1109.0199 [hep-ph]}].

\bibitem{arXiv:0808.3372}
  C.~S.~Fischer and R.~Williams,
  Phys.\ Rev.\ D\ {\bf 78} (2008) 074006
  [{\tt arXiv:0808.3372 [hep-ph]}].

\bibitem{arXiv:0905.2291}
  C.~S.~Fischer and R.~Williams,
  Phys.\ Rev.\ Lett.\ \ {\bf 103} (2009) 122001
  [{\tt arXiv:0905.2291 [hep-ph]}].

\bibitem{arXiv:0903.5461}
  L.~Chang and C.~D.~Roberts,
  Phys.\ Rev.\ Lett.\ \ {\bf 103} (2009) 081601
  [{\tt arXiv:0903.5461 [nucl-th]}].

\bibitem{arXiv:0811.1752}
  W.~Plessas and T.~Melde,
  AIP Conf.\ Proc.\ \ {\bf 1056} (2008) 15
  [{\tt arXiv:0811.1752 [hep-ph]}].

\bibitem{nucl-th/9708029}
  P.~Maris and C.~D.~Roberts,
  Phys.\ Rev.\ C\ {\bf 56} (1997) 3369
  [{\tt nucl-th/9708029}].

\bibitem{nucl-th/9905056}
  P.~Maris and P.~C.~Tandy,
  Phys.\ Rev.\ C\ {\bf 60} (1999) 055214
  [{\tt nucl-th/9905056}].

\bibitem{arXiv:0804.3478}
  R.~Alkofer, C.~S.~Fischer and R.~Williams,
  Eur.\ Phys.\ J.\ A\ {\bf 38} (2008) 53
  [{\tt arXiv:0804.3478 [hep-ph]}].

\bibitem{hep-ph/0309077}
  R.~Alkofer, W.~Detmold, C.~S.~Fischer and P.~Maris,
  Phys.\ Rev.\ D\ {\bf 70} (2004) 014014
  [{\tt hep-ph/0309077}].

\bibitem{arXiv:0804.3042}
  R.~Alkofer, C.~S.~Fischer, F.~J.~Llanes-Estrada and K.~Schwenzer,
  Annals Phys.\ \ {\bf 324} (2009) 106
  [{\tt arXiv:0804.3042 [hep-ph]}].

\bibitem{hep-lat/0501021}
  T.~-W.~Chiu and T.~-H.~Hsieh,
  Nucl.\ Phys.\ A\ {\bf 755} (2005) 471
  [{\tt hep-lat/0501021}].

\bibitem{arXiv:1010.0889}
  R.~Lewis,
  AIP Conf.\ Proc.\ \ {\bf 1374} (2011) 581
  [{\tt arXiv:1010.0889 [hep-lat]}].

\bibitem{arXiv:1111.7087}
  F.~J.~Llanes-Estrada, O.~I.~Pavlova and R.~Williams,
 [{\tt arXiv:1111.7087 [hep-ph]}].

\bibitem{arXiv:1007.3901}
  M.~Blank, A.~Krassnigg and A.~Maas,
  Phys.\ Rev.\ D\ {\bf 83} (2011) 034020
 [{\tt arXiv:1007.3901 [hep-ph]}].

\bibitem{arXiv:1108.0603}
  S.-X.~Qin, L.~Chang, Y.-X.~Liu, C.~D.~Roberts and D.~J.~Wilson,
  Phys.\ Rev.\ C\ {\bf 84} (2011) 042202
 [{\tt arXiv:1108.0603 [nucl-th]}].

\end{thebibliography}
\end{document}